\documentclass[12pt,preprint]{aastex}

\begin{document}
\title{Explaining the mysterious age gap of globular clusters in the Large Magellanic Cloud}

\author{Kenji Bekki\altaffilmark{1} and Warrick J. Couch \altaffilmark{1}} 

\author{Michael A. Beasley,\altaffilmark{2} Duncan  A. Forbes\altaffilmark{2}}

\author{Masashi Chiba\altaffilmark{3}} 

\and

\author{Gary S. Da Costa\altaffilmark{4}}

\altaffiltext{1}{School of Physics, University of New South Wales, 
Sydney 2052, Australia, bekki@bat.phys.unsw.edu.au}
\altaffiltext{2}{ Centre for Astrophysics \& Supercomputing, 
Swinburne University of Technology, 
Hawthorn, VIC,  3122, Australia, mbeasley@mania.physics.swin.edu.au}
\altaffiltext{3}{Astronomical Institute, 
Tohoku University, Sendai, 980-8578, Japan, chibams@crocus.ocn.ne.jp}
\altaffiltext{4}{Research School of Astronomy \& Astrophysics,
Institute of Advanced Studies,  Australian National University,
Cotter Road, Weston Creek,    ACT 2611, Australia, gdc@mso.anu.edu.au}

\begin{abstract}

The Large Magellanic Cloud (LMC) has a unique cluster formation history
in that nearly all of its globular clusters were formed either $\sim$ 13
Gyr ago or less than $\sim$ 3 Gyr ago.
It is not clear what physical mechanism is responsible for the most
recent cluster formation episode and thus the mysterious age gap
between the LMC clusters.
We first present results of gas dynamical
N-body simulations
of the evolution of the LMC in the context of its Galactic orbit
and interactions with the SMC, paying special attention to the effect
of tidal forces.
We find that
the first close encounter
between the LMC and the Small Magellanic Cloud (SMC) about 4 Gyr ago
was the beginning of a period of strong tidal interaction which
likely induced dramatic gas cloud collisions, leading to an enhancement
of the formation of globular clusters which has been sustained by
strong tidal interactions to the present day.
The tidal interaction results in the formation of a barred, elliptical,
thick disk in the LMC.
The model also predicts the presence
of a large, diffuse stellar stream circling the Galaxy, which originated
from the LMC.

\end{abstract}


\section{Introduction}

Tidal interactions between galaxies are suggested, both from observations
and theory, to dramatically change the formation rates of field stars and
globular clusters due to the tidal compression of gas clouds and their
efficient conversion into stars 
(Kennicutt 1998; Ashman \& Zepf 1992;  Noguchi \& Ishibashi 1986;
Bekki \& Couch 2001; Bekki et al. 2002). 
The Large and Small Magellanic
Clouds (LMC and SMC), which are the nearest pair of interacting galaxies
in our vicinity, have served as unique laboratories for studying the
interplay between galactic dynamical evolution and star formation
activity in galaxies (Westerlund 1997; van den Bergh 2000). 
A major curiosity in this context is
that globular cluster formation -- which is considered to be a special
mode of star formation 
(Harris \&  Pudritz 1994; Elmegreen \&  Efremov 1997)
-- is observed to be ongoing in the
Clouds but not in the Galaxy. Clearly the physical conditions
required for cluster formation currently exist in the Clouds
but not in the  Galaxy (Westerlund 1997; van den Bergh 2000). 
Moreover, the age distribution of the LMC
clusters shows a gap extending from 3 to 13 Gyr--
with only one cluster in this age range--
suggesting a second epoch of cluster formation started abruptly in the LMC
about 3 Gyr ago 
(e.g., Da Costa 1991;  Geisler et al. 1997; Rich et al. 2001;  Piatti et al. 2002). 
The origin of the ``age gap'', which has not
been observed for the SMC, has remained elusive so far
(e.g., Da Costa 1991).
The purpose of this Letter is to provide
a new explanation for the origin of the mysterious age gap of the LMC's
globular cluster system. 
The present study also tries to explain the latest observational results of 
the LMC's structural and kinematical properties
(e.g., Alves \& Nelson 2000; van der Marel et al. 2002).

\section{The model}

We present here a new numerical model of the dynamical evolution and the
star formation history of the LMC as it orbits the Galaxy and is
gravitationally influenced both by the Galaxy and the SMC, in an attempt
to explain the origin of the age gap. We first determine the most
plausible and realistic orbits of the Clouds and then investigate the
evolution of the LMC using fully self-consistent N-body models. In
determining the orbits, we adopt the same numerical method as those in
previous studies 
(Murai \&  Fujimoto 1980;  Gardiner et al. 1994; Gardiner \&  Noguchi 1996), 
in which the equations of motion of the clouds
are integrated backwards in time, from the present epoch until $\sim$ 9
Gyrs ago. Figure 1 shows the best orbit model for the Clouds calculated in this way,
where it can be seen they experience a very close tidal encounter (at a
pericenter distance of 6.4 kpc) around 3.6 Gyr ago (T$\sim -3.6$)  and
then become dynamically coupled. 
If dynamical friction between the Clouds and the mass
of the LMC ($10^{10}$ $\rm M_{\odot}$) are considered 
-- something which has not been done in previous
studies (e.g., Gardiner et al. 1994),
then the present-day binary nature of the Clouds cannot 
have existed for a Hubble time: They must have been
separate objects in the past.

Using gas-dynamical  N-body simulations (Bekki et al. 2004),
we investigate dynamical evolution and star formation 
histories of the LMC for the best orbit model.
The LMC is modeled here as a bulgeless,
gas-rich disk embedded in a massive dark matter halo with a mass 2.3
times larger than the disk mass, whereas the SMC is treated as a
point mass with the mass of $3 \times 10^{9}$ $\rm M_{\odot}$. The exponential old
disk of the LMC is taken 
to have a scale length of 2.6\,kpc (Kinman et al. 1991) and
a maximum circular velocity ($V_{\rm m}$) of 71\,km\,s$^{-1}$,
consistent with recent observations (van der Marel et al. 2002). 
The gas disk is represented by
a collection of discrete gas clouds that follow the observed mass-size
relationship (Larson 1981). Each pair of two overlapping gas clouds
collides with the same restitution coefficient of 0.5 
(Hausman \& Roberts 1984).

Globular cluster formation in the present model is discriminated from
field star formation as follows:  The gas is converted into field stars
according to the Schmidt law with the observed threshold
gas density (Kennicutt 1998).  
We use the cluster formation criteria derived by
previous analytical works (e.g., Kumai et al. 1993)
and hydrodynamical simulations  with variously different parameters of
cloud-cloud collisions on a 1-100pc scale (Bekki et al. 2004) 
in order to model globular
cluster formation. A gas particle is converted into a cluster if it
collides with other high velocity gas (ranging from 30\,km\,s$^{-1}$
to 100\,km\,s$^{-1}$) and having an impact parameter (normalized to the
cloud radius) less than 0.25.  
This model is strongly supported by recent observations
(e.g., Zhang et al. 2001) that have revealed that
there is a tendency for young clusters to be found
in gaseous regions with higher velocity dispersion,
where cloud-cloud collisions are highly likely. 
Chemical evolution of the gas with an
assumed effective chemical yield of 0.005 is also incorporated in the
simulations. 
Our numerical results on field star formation histories 
in the LMC do not depend strongly on
the above parameters of gas dynamics (e.g., gas mass fraction)
and star formation (the exponent of the Schmidt law) for a reasonable
set of parameters.

In the present paper, we show only the results of the best model
in which the Clouds first came together  about 4 Gyr ago. 
It should be here stressed that this conjunction is not 
a unique characteristic of the best model: About 95 \% of the orbital
models have a disintegration of the LMC-SMC binary
-- recall
the integrations are done backwards in time -- occurring within the last
4 Gyr for  models with the orbital parameters consistent
with observations. 
We have investigated orbital evolution of the Clouds for
$4 \times 10^8$  models with
the current space velocities
($V_{\rm X}$,$V_{\rm Y}$,$V_{\rm Z}$)
ranging from $-320$ km s$^{-1}$ to $320$ km s$^{-1}$ 
for the Clouds and thereby derived the epoch
when the Clouds become separated.
We have confirmed that (1) it is highly unlikely
that  the Clouds were a  binary system  more than 6 Gyr ago,
and (2) they most likely came together around 4 Gyr ago.
Thus our main conclusions on the epoch of
the commencement of cluster
formation in the LMC do not depend 
so strongly on orbital parameters.

\placefigure{fig-1}
\placefigure{fig-2}

\section{Results}

As seen in Figure 2, the combined tidal effects of the SMC and the Galaxy
distorts the LMC disk (T=$-6.6$ and $-3.3$ Gyr) to form compact stellar
and gaseous bars in its center after a few  pericenter passages
(T=$-3.3$\,Gyr). Consequently, young stars are formed repeatedly in the
tidally compressed high-density gaseous regions, in particular the
central bar, where the star formation rate is at a maximum at
0.38 $\rm M_{\odot}$ yr$^{-1}$. The tidal interaction significantly
increase the degree of random motion of the gas clouds so that they
collide much more frequently with one another. However, the cloud-cloud
collisions with moderately high speed (between 30 and 100 km s$^{-1}$)
and small impact parameter ($<$ 0.25), required for cluster formation,
do not occur until the LMC begins to interact violently with the
SMC when the two are less than 10\,kpc apart (T=$-3.6$\,Gyr). The
dynamical evolution over the last $\sim$~3\,Gyr results in about
0.5\% of the initial gas cloud mass being converted into clusters.
The new globular cluster system has a flattened, `disky' spatial
distribution with nearly all clusters within the central
$\sim$ 3 kpc and confined within $\sim$ 1 kpc from the LMC disk plane
(i.e., an indication of a structure similar to the Galactic thick disk).
They are supported mainly by rotation with 
the central velocity dispersion of only 20 km s$^{-1}$ and
with a fairly asymmetric radial dispersion profile caused by
recent tidal interaction with the Galaxy and the SMC.
The derived structure and kinematics of clusters
are broadly consistent with the observed ones for intermediate-age
clusters (Schommer et al. 1992).
Due to the rapid
chemical enrichment associated with this star formation activity,
the mean metallicity of clusters increase from [Fe/H] = $-0.91$ to
$-0.33$ over this period. Such a recently formed cluster population
consequently has a narrower age distribution 
than the new field stars (Figure 3). Thus the origin of the age gap
and in particular the trigger for the most recent episode of globular
cluster formation in the LMC may well be related to the commencement of
strong tidal interactions between the LMC, the SMC and the Galaxy.

The tidal interaction between the three may also play an important role
in changing the structural and kinematical properties of the LMC.
Because of the strong tidal perturbation, the initial thin, non-barred
disk is dynamically heated to form a thick, barred disk with an outer
stellar warp. Furthermore, $\sim 17$\% of the stars in the disk
are spatially redistributed to form an outer stellar halo with a
velocity dispersion of $\sim 40$\,km\,s$^{-1}$ at a distance of 7.5\,kpc
from the LMC center.
The kinematically hot stellar halo is dominated by
old stars with the fraction of new field stars equal to only 2\%,
because most halo stars originate from the outer part of the initial thin disk.
In contrast, $\sim 56$\% of the stars within the central disk are new
field stars formed from the triggered star formation. For example, our
simulations indicate that the half mass radius is $\sim 2.1$\,kpc for the old
field populations and $\sim 0.9$\,kpc for the new populations
(i.e., the different structure of the different aged populations).
The stellar kinematics along the major axis shows the central velocity
dispersion, ${\sigma}_{0}$, to be $\sim 30$\,km\,s$^{-1}$ with $V_{\rm
m}/{\sigma}_{0}$ of $\sim$ 2.3.
These results are broadly consistent with observations by
van der Marel et al. (2002).
These also suggest that the LMC stellar halo 
has a significantly larger  fraction of 
young, relatively metal-rich ([Fe/H] $<$  $-0.3$)
stellar populations compared with the Galactic 
stellar halo dominated by very old, metal-poor 
([Fe/H] $\sim$  $-1.6$) stars.

The origin of the field stars formed in recent star formation activity in
the LMC disk is one of the most important problems associated with the
formation of the LMC 
(e.g., Butcher 1997;  Olszewski et al. 1996; Gallagher et al. 1996).
Our evolutionary model strongly suggests
that such recent star formation is due to the repetitive tidal interaction
between the LMC/SMC and the Galaxy. 
Field star formation is more sensitive to  tidal perturbation
than cluster formation so that enhancement in field star formation
can occur from the early evolutionary stage of the LMC (i.e., $6-7$ Gyr ago).
This can clearly explain why ``the age gap'' can be more clearly
observed in cluster populations than in field ones. 
Chemical enrichment associated with
the tidally triggered star formation provides a natural explanation for
the observed rapid increase of metallicity both in clusters and in field
stars (Da Costa 1991; Dopita et al. 1997).

What then are the fossil records of any past dynamical interaction?
The tidal radius of the LMC, within which its stars are gravitationally
bound, is linearly proportional to the pericenter of the orbit for the
fixed masses of the LMC and the Galaxy (Gardiner \&  Noguchi 1996). 
Therefore, as the
pericenter of the LMC orbit decreases as a result of dynamical
friction, it loses an increasingly larger number of stars.
Figure 4 shows that the stripped stars in the simulation form a great
circle - a relic stream in which the stars are diffusely and
inhomogeneously distributed along the LMC's orbit in the Galactic halo
region. Such a stream has
been predicted in previous studies (Weinberg 2000), 
but for the first time our
simulations  provide quite clear predictions on the spatial
distribution and radial velocities of stars within the stream. Ongoing
and future surveys for stellar substructure in the Galactic halo 
(Majewski  et al. 2000) will
allow these predictions to be tested. Indeed preliminary observational
results have already suggested that the coherent radial velocity
structures amongst giant stars in fields encircling the Clouds are due
to tidal debris that originated from the LMC (Majewski  et al. 2000).

\placefigure{fig-3}
\placefigure{fig-4}

\section{Discussion and conclusion}

How then can the model explain the observed different
age distributions of clusters
between the LMC and the SMC (Piatti et al. 2002)? We propose 
that such a difference can be understood in terms of the differences in 
the birthplaces and initial masses between the two. The LMC was formed as 
a relatively low surface brightness galaxy, being more distant 
($\sim 150$\,kpc, corresponding to the apocenter of its early orbit) 
from the Galaxy so that the Galactic tidal field alone could not trigger 
cluster formation efficiently until it first encounters with the SMC.
In contrast, the less massive SMC, which is therefore more susceptible to 
the Galactic tide, was born less distant ($\sim 100$\,kpc) from the 
Galaxy, and thus influenced by the Galaxy strongly enough to form  
globular clusters from the early evolutionary stage (several to 10\,Gyrs 
ago). Thus the difference in cluster formation histories could
reflect the fact that the LMC/SMC were formed as different entities
rather than as a binary protogalaxy.

Tidal interaction between the Clouds and the Galaxy has long been
considered to be a physical mechanism responsible for the formation
of the Magellanic stream and the evolution of the SMC
(Murai \& Fujimoto 1980; Lin \& Lynden-Bell 1982).
We have demonstrated that such interaction can also cause dramatic changes
not only in the formation history of field stars and globular clusters
but also in the LMC's structure and kinematics. Our model predicts that
if globular cluster formation is associated with a dramatic increase in
the random motions in the LMC's interstellar medium, then the age-,
metallicity- and spatial-distributions of these systems will be very
much dependent on the orbital evolution of the Clouds. This can be
inferred observationally from radial velocity and proper motion
measurements of their stars. Future proper motion measurements with
$\sim 10$\,micro-arcsec accuracy (Perryman et al. 2001), coupled with numerical
simulations for different yet reasonable orbits of the Clouds,
will therefore enable us to obtain an integrated and systematic
understanding of the formation of the LMC clusters.

\acknowledgments

We are  grateful to the anonymous referee for valuable comments,
which contribute to improve the present paper.
K.B.,  W.J.C., and G. Da C.  acknowledge financial support from the Australian
Research Council (ARC) throughout the course of this work. The
numerical simulations reported here were carried out on GRAPE
systems kindly made available by the Astronomical Data Analysis
Center (ADAC) at National Astronomical Observatory of Japan (NAOJ).\\

\clearpage


\figcaption{
Schematic view of the Galaxy and the Magellanic Clouds (upper)
and the orbital evolution of the Clouds (lower).
The current Galactic coordinate $(b,l)$, where $l$ and $b$ are
the Galactic longitude and latitude, respectively, is $(-32.89, 280.46)$
for the LMC and $(-44.30,302.79)$ for the SMC, and accordingly
the current positions $(X,Y,Z)$ in units of kpc in the figure
are $(-1.0,-40.8,-26.8)$  for the LMC
and $(13.6,-34.3,-39.8)$ for the SMC.
The current distance and the Galactocentric radial velocity of the LMC (SMC)
is 80 (7) km s$^{-1}$. The total masses of the LMC and the SMC are
1.0 $\times$ $10^{10}$ $M_{\odot}$ and 3.0 $\times$ $10^9$ $M_{\odot}$,
respectively. These values were based on those derived from
the latest observations 
(e.g., Kroupa \&  Bastian 1997; van der Marel et al 2002)
and consistent with previous simulations (e.g., Gardiner \&  Noguchi 1996).
In the Clouds' orbital calculations with the backward integration scheme
(Gardiner et al. 1994),
we assume that the current space velocities $(V_{\rm x}, V_{\rm y}, V_{\rm z})$
(or  $(U,V,W)$) in units of km s$^{-1}$
are $(-5,-225,194)$ and $(40,-185,171)$.
We chose these values
firstly because the Magellanic stream can be self-consistently
reproduced in previous models for these values,
secondly because they are  broadly consistent with
the latest proper motion data.
The Galactic gravitational potential ${\Phi}_{\rm G}$ is represented by
${\Phi}_{\rm G}=-{V_{0}}^2 {\rm ln} r$, where $V_{0}$ and $r$ are
the constant rotational velocity (220 km s$^{-1}$ in this study) and the distance
from the Galactic center (G.C.).
A Plummer potential is adopted for the LMC/SMC with a softening length
of 3(2)\,kpc for the LMC (SMC). For the adopted velocities and positions,
the orbits of the Clouds are nearly polar with the Clouds leading the
Magellanic stream. Negative values of the time, $T$, represent the past, with
$T=0$ corresponding to the present epoch. Note that the LMC-SMC distance
(shown as the {\it thick magenta line}) remains very small
($<40$\,kpc) over the last 4\,Gyr ($T >$  $-4$ Gyr), due to the dynamical
coupling of the LMC and SMC.
\label{fig-1}}

\figcaption{
Morphological evolution of the LMC
for the last $\sim 9$\,Gyr (upper)
and the time evolution of the star formation rate
and the number of cloud-cloud collisions  (lower).
Magenta and blue represent  ``old'' stellar
and gaseous components and  ``new'' stars, respectively.
Here ``old'' and ``new'' stars mean field stars initially in the LMC disk
and those newly formed from gas, respectively. The total number of
particles used in the simulation is $1.4 \times 10^5$.
Star formation rates are measured in units of $\rm M_{\odot}$ yr$^{-1}$.
The number of high-velocity cloud-cloud collisions with relative
velocities ranging from 30 to 100\,km\,s$^{-1}$ (required for
globular cluster formation) are counted at each time
in the lower panel. Note that a bar composed of new stars is formed
through repetitive tidal interaction until $T=$-3.3\,Gyr.
Note also that the cloud-cloud collision rate becomes very high
after the first close encounter (at $T\sim$-3.6\,Gyr) of the LMC/SMC.
The star formation rate also becomes moderately high
during close tidal encounters.
\label{fig-2}}

\figcaption{
Final age
distribution of
newly formed field stars and star clusters.
New field stars and  clusters
are represented by dotted and solid lines, respectively.
The six epochs of the LMC-SMC pericenter passage
are indicated by thick arrows in the upper panel for comparison.
All clusters have ages younger than $\sim$ 3.3\,Gyrs
whereas the field star population show a wide distribution
of ages.  This result reflects the fact that
the field star formation is sensitive to local gas density whereas
the cluster formation can happen only when random motion
of gas in the LMC becomes significantly large.
\label{fig-3}}

\figcaption{
The distribution of old stars stripped from the
LMC disk in an Aitoff projection.
Only stars stripped from the LMC
(corresponding to 17\% of the initial LMC old stars)
for the last $\sim 9$\,Gyr are plotted.
The locations of the LMC (larger filled circles) and the SMC (smaller)
at the present (blue) and 9\,Gyr ago (red) are shown for comparison.
An appreciable crowding of stars can be seen around
$60\le l\le 120$ and $-30\le b\le 30$ in the relic stellar stream.
The distances of stars from the Galactic center
range from 25 to 250\,kpc with a mean of 91\,kpc.
The Galactocentric radial velocities range from
-213 to 223\,km\,s$^{-1}$ with a mean of -4\,km\,s$^{-1}$.
\label{fig-4}}

\newpage
\plotone{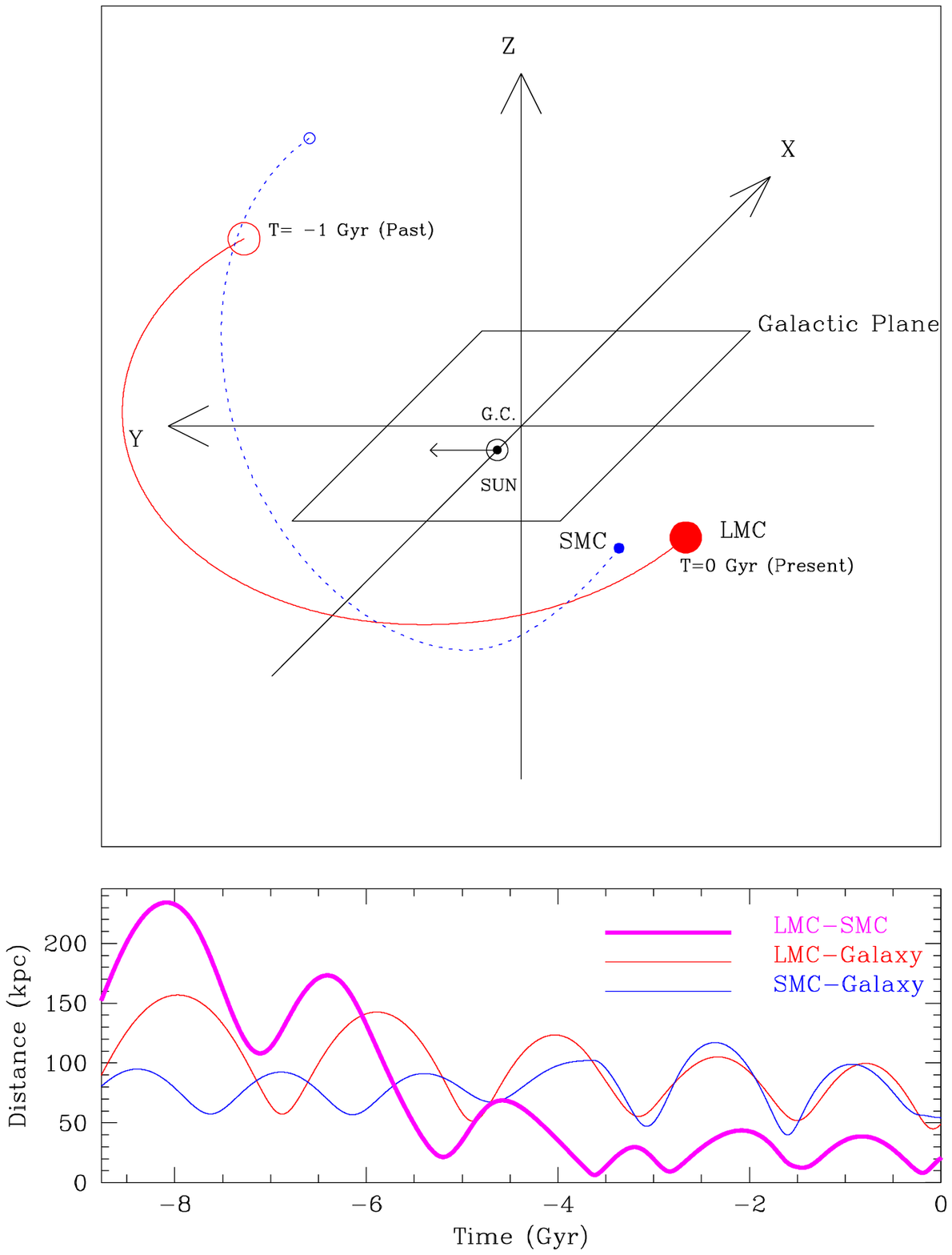}
\newpage
\epsscale{0.8}
\plotone{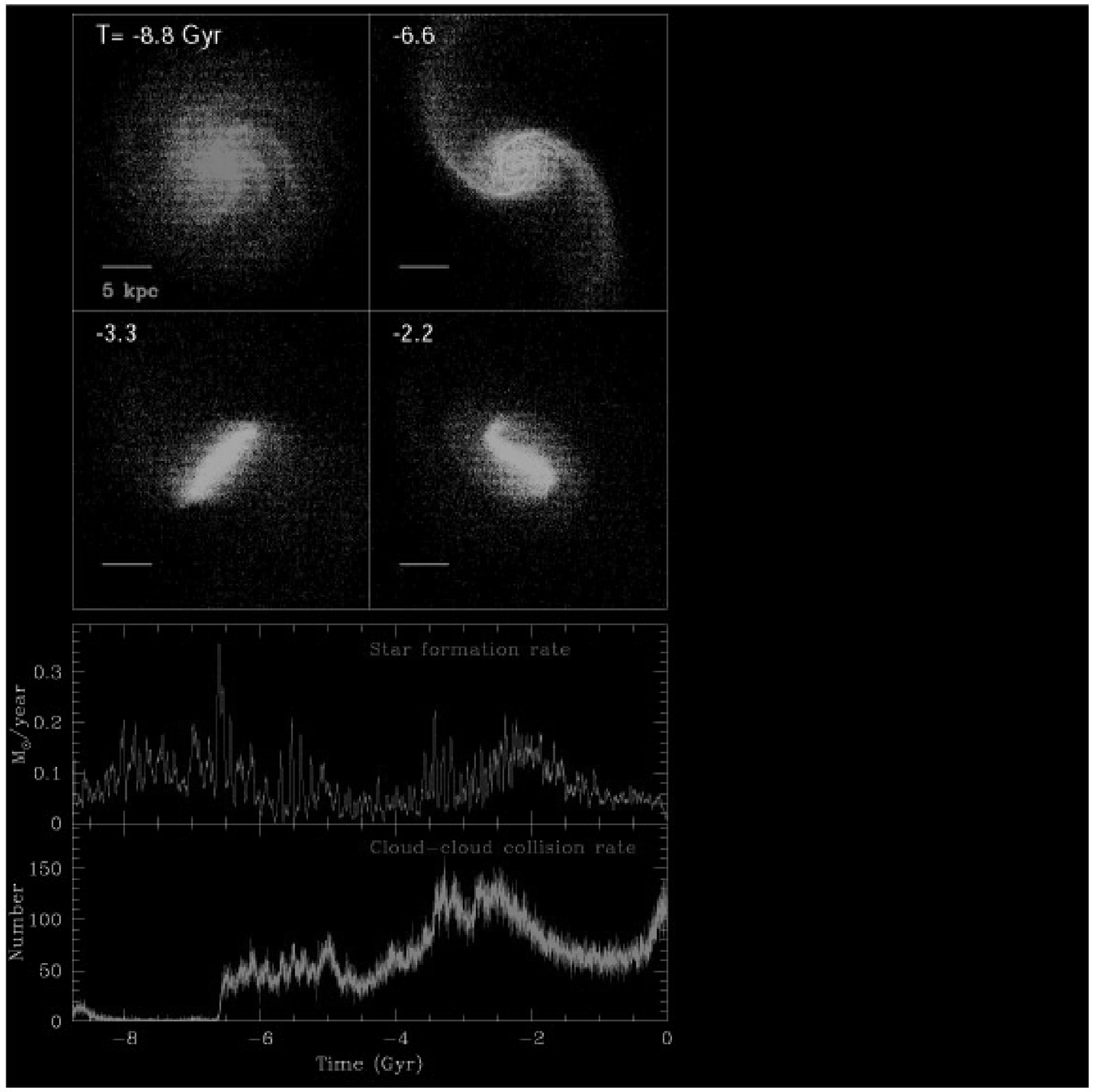}
\newpage
\epsscale{1}
\plotone{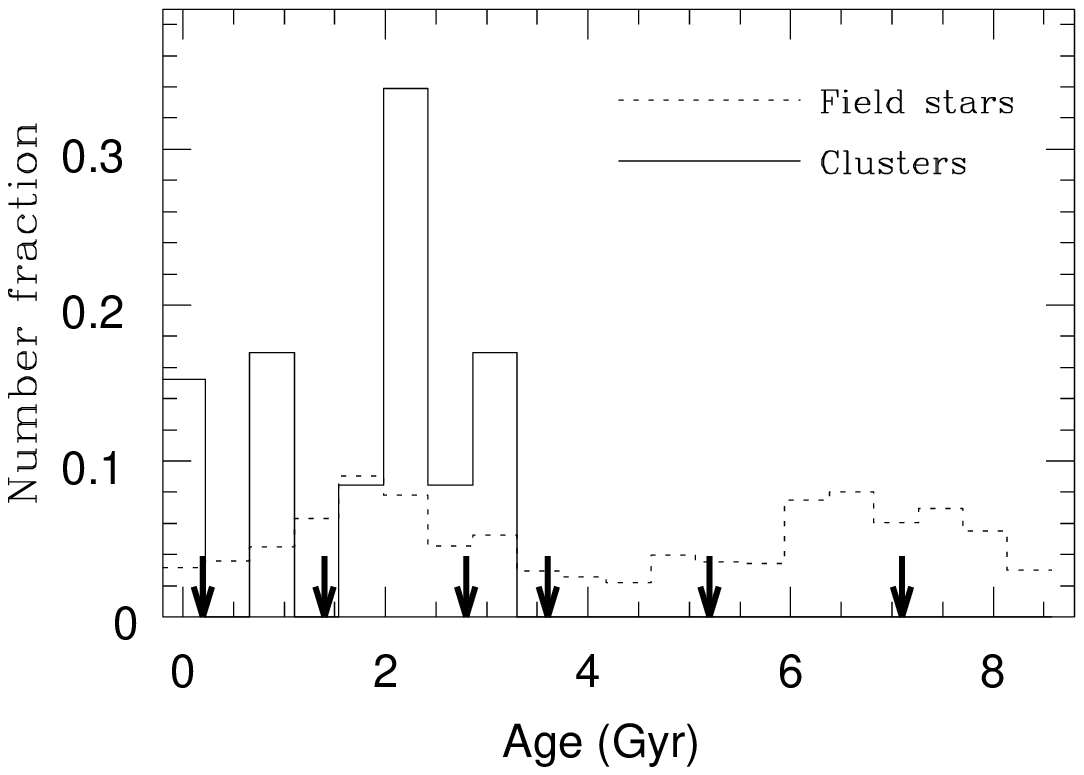}
\newpage
\plotone{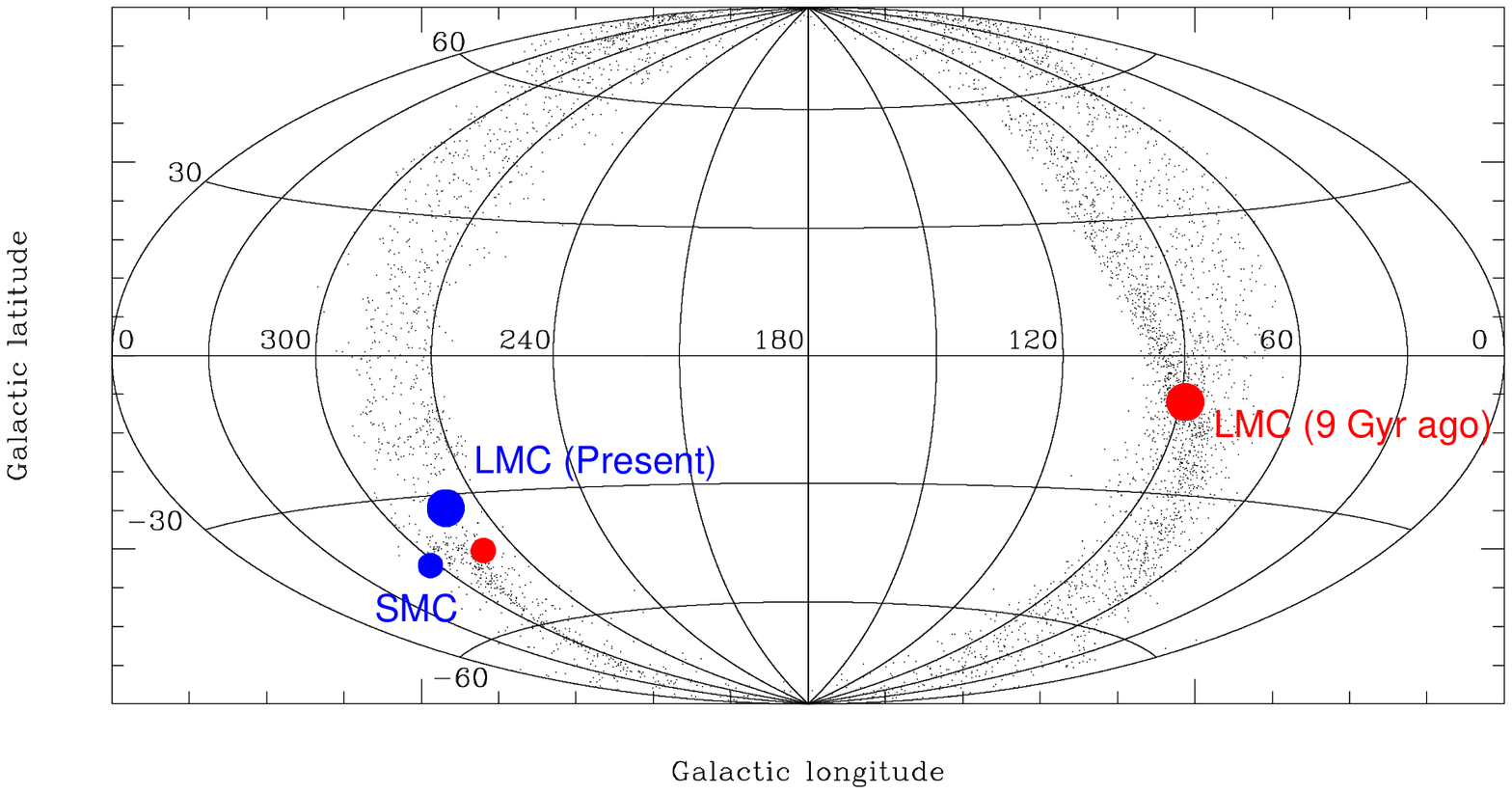}

\end{document}